\pdfoutput=1
\documentclass[sn-standardnature]{sn-jnl}

\usepackage{diagbox}
\usepackage[version=4]{mhchem}
\usepackage{gensymb}
\usepackage{lineno}

\jyear{2023}%

\theoremstyle{thmstyleone}%
\theoremstyle{thmstyletwo}%

\theoremstyle{thmstylethree}%

\begin{document}

\title[Muon Beam Emittance Reduction by Ionization Cooling]{Transverse Emittance Reduction in Muon Beams by Ionization Cooling}

\subtitle{MICE Collaboration}



\author[1]{\fnm{M.} \sur{Bogomilov}} 
\author[1]{\fnm{R.} \sur{Tsenov}} 
\author[1]{\fnm{G.} \sur{Vankova-Kirilova}}

\author[2]{\fnm{Y.~P.} \sur{Song}} 
\author[2]{\fnm{J.~Y.} \sur{Tang}} 

\author[3]{\fnm{Z.~H.} \sur{Li}} 

\author[4]{\fnm{R.} \sur{Bertoni}} 
\author[4]{\fnm{M.} \sur{Bonesini}} 
\author[4]{\fnm{F.} \sur{Chignoli}}
\author[4]{\fnm{R.} \sur{Mazza}}

% ...

\affil[1]{\orgdiv{Department of Atomic Physics}, \orgname{St.~Kliment Ohridski University of Sofia}, \orgaddress{\street{5 James Bourchier Blvd}, \city{Sofia}, \country{Bulgaria}}}

\affil[2]{\orgdiv{Institute of High Energy Physics}, \orgname{Chinese Academy of Sciences}, \orgaddress{\street{19 Yuquan Rd, Shijingshan District}, \city{Beijing}, \country{China}}}

\affil[3]{\orgname{Sichuan University}, \orgaddress{\street{252 Shuncheng St}, \city{Chengdu}, \country{China}}}

\affil[4]{\orgdiv{Sezione INFN Milano Bicocca}, \orgname{Dipartimento di Fisica G.~Occhialini}, \orgaddress{\street{Piazza della Scienza 3}, \city{Milano}, \country{Italy}}}

\affil[5]{\orgdiv{Sezione INFN Pavia and Dipartimento di Fisica}, \orgname{Universit\`{a} di Pavia}, \orgaddress{\street{Via Agostino Bassi 6}, \city{Pavia}, \country{Italy}}}

% ...
\affil[19]{\orgdiv{School of Physics and Astronomy, Kelvin Building}, \orgname{University of Glasgow}, \orgaddress{\city{Glasgow}, \country{UK}}}

% ...


\abstract{


Accelerated muon beams have been considered for next-generation studies of high-energy lepton-antilepton collisions and neutrino oscillations. However, high-brightness muon beams have not yet been produced. The main challenge for muon acceleration and storage stems from the large phase-space volume occupied by the beam, derived from the muon production mechanism through the decay of pions from proton collisions. Ionization cooling is the technique proposed to decrease the muon beam phase-space volume. Here we demonstrate a clear signal of ionization cooling through the observation of transverse emittance reduction in beams that traverse lithium hydride or liquid hydrogen absorbers in the Muon Ionization Cooling Experiment (MICE). The measurement is well reproduced by the simulation of the experiment and the theoretical model. The results shown here represent a substantial advance towards the realization of muon-based facilities that could operate at the energy and intensity frontiers.
}

\maketitle

\section*{Muon-beam facilities}\label{intro}

Muon accelerators are considered potential enablers of fundamental particle physics studies at the energy and intensity frontiers. Such machines have great potential to provide multi-TeV lepton--antilepton collisions at a muon collider \cite{neuffer1996progress,muoncollider1,palmer2014muon} or act as sources of intense neutrino beams with well-characterized fluxes and energy spectra at a neutrino factory \cite{geer1998neutrino,DERUJULA199921,PhysRevSTAB.17.121002}.

The benefit of using muons in circular storage rings arises from their fundamental nature and their mass which is 207 times that of electrons. As elementary particles, colliding muons offer the entire centre-of-mass energy to the production of short-distance reactions. This is an advantage over proton-proton colliders, such as the Large Hadron Collider \cite{evans2007lhc}, where the colliding proton constituents each carry only a fraction of the proton energy. In comparison with the electron, the larger muon mass leads to a dramatic reduction in synchrotron radiation losses, which scale as $1/m^4$. In addition, the spread in the effective centre-of-mass energy induced by beamstrahlung \cite{noble1987beamstrahlung}, the emission of radiation resulting from the interaction of a charged particle beam with the electric field produced by an incoming beam, is substantially lower for muons. Thus, a muon collider could achieve multi-TeV and precise centre-of-mass energies with a considerably smaller facility than an electron-positron collider such as the proposed electron-positron variant of the Future Circular Collider \cite{fcc2019}, the Circular Electron-Positron Collider \cite{cepc2018cepc}, the International Linear Collider \cite{behnke2013ilc} or the Compact Linear Collider \cite{clic2018compact}.

The primary challenges in building a muon collider facility stem from the difficulty of producing intense muon bunches with a small phase-space volume, as well as the short muon lifetime (2.2$\,\mathrm{\mu}$s at rest). A proton-driver scheme is currently the most attractive option due to its potential to generate intense muon beams. An alternative, positron-driven muon source has been proposed and is under conceptual study \cite{antonelli2016novel}. In the proton-driver scheme, an intense proton beam impinges on a target to produce a secondary beam composed primarily of pions and kaons. The pions and kaons decay into muons to create a tertiary muon beam. The resulting muon beam occupies a large phase-space volume, which must be reduced (cooled) to allow efficient acceleration and sufficient flux and luminosity. The muon capture, cooling and acceleration must be executed on a time scale comparable with the muon lifetime.

Traditional cooling techniques such as stochastic cooling \cite{mohl1980physics}, electron cooling \cite{parkhomchuk2000electron} or synchrotron radiation cooling \cite{kolomenski1956effect} are impractical as the amount of time required to cool the beam adequately greatly exceeds the muon lifetime. Ionization cooling is the proposed technique by which the muon beam phase-space volume can be compressed sufficiently before significant decay losses occur \cite{skrinskii1981methods,neuffer}. Ionization cooling occurs when a muon beam passes through a material, known as the absorber, and loses both transverse and longitudinal momentum by ionizing atoms. The longitudinal momentum can be restored using radiofrequency (RF) accelerating cavities. The process can be repeated to achieve sufficient cooling within a suitable time frame \cite{stratakis2015rectilinear}.


The Muon Ionization Cooling Experiment (MICE; \href{http://mice.iit.edu}{http://mice.iit.edu}) was designed to provide the first demonstration of ionization cooling by measuring a reduction in the muon beam transverse emittance after the beam has passed through an absorber. A first analysis conducted by the MICE collaboration has demonstrated an unambiguous cooling signal by observing an increase in the phase-space density in the core of the beam upon passage through an absorber \cite{cooldemo}. Here we present the quantification of the ionization cooling signal by measuring the change in the beam's normalized transverse emittance, which is a central figure of merit in accelerator physics. A novel beam sampling procedure is employed to improve the measurement of the cooling performance by selecting muon subsamples with optimal beam optics properties in the experimental apparatus. This beam sampling enables the probing of the cooling signal in beams with lower input emittances than those studied in \cite{cooldemo} and facilitates a comparison between the measurement and the theoretical model of ionization cooling.




\section*{Ionization cooling}\label{beamemitt}


The normalized root-mean-square (R.M.S.) emittance is a measure of the volume occupied by the beam in phase-space. It is a commonly used quantity in accelerator physics that describes the spatial and dynamical extent of the beam, and it is a constant of motion under linear beam optics. This work focuses on the four-dimensional phase-space transverse to the beam propagation axis. The MICE coordinate system is defined such that the beam travels along the $z$-axis, and the state vector of a particle in transverse phase-space is given by $\textbf{u} = (x, p_x, y, p_y)$. Here $x$ and $y$ are the position coordinates and $p_x$ and $p_y$ are the momentum coordinates. The normalized transverse R.M.S. emittance is defined as \cite{wiedemann2015particle}

\begin{equation}
 \varepsilon_{\perp} = \frac{1}{m_{\mu}c}\left\vert\boldsymbol{\Sigma}_{\perp}\right\vert^{\frac{1}{4}},
    \label{eq:em4d}
\end{equation}
where $m_{\mu}$ is the muon mass and $\left\vert\Sigma_{\perp}\right\vert$ is the determinant of the beam covariance matrix. The covariance matrix elements are calculated as $\Sigma_{\perp, ij} = \langle u_iu_j \rangle -\langle u_i \rangle \langle u_j \rangle$.

The impact of ionization cooling on a beam crossing an absorber is best described through the rate of change of the normalized transverse R.M.S. emittance, which is approximately equal to \cite{neuffer,penn2000beam,pbj2022}

\begin{equation}\label{eq:cooleqn}
    \frac{d\varepsilon_{\perp}}{dz} \simeq - \frac{\varepsilon_{\perp}}{\beta^2 E_{\mu}} \left\vert\frac{dE_{\mu}}{dz}\right\vert + \frac{\beta_{\perp} (13.6\,\mathrm{MeV/c})^2}{2\beta^3E_{\mu}m_{\mu}X_{0}},
\end{equation}
where $\beta c$ is the muon velocity, $E_{\mu}$ the muon energy, $\left\vert dE_{\mu}/dz \right\vert$ the average rate of energy loss per unit path length, $X_0$ the radiation length of the absorber material, and $\beta_\perp$ the beam transverse betatron function at the absorber defined as $\beta_\perp = \frac{\langle x^2 \rangle + \langle y^2\rangle}{2m_\mu c\varepsilon_\perp}\langle p_z \rangle\,.$
The emittance reduction (cooling) due to ionization energy loss is expressed through the first term. The second term represents emittance growth (heating) due to multiple Coulomb scattering by the atomic nuclei, which increases the angular spread of the beam. MICE recently measured scattering in lithium hydride and observed good agreement with the GEANT4 model \cite{MICE:2022ffq}.

The cooling is influenced by both the beam properties and the absorber material. Heating is weaker for beams with lower transverse betatron function at the absorber. This can be achieved by using superconducting solenoids that provide strong symmetrical focusing in the transverse plane. The absorber material affects both terms in the equation, and optimal cooling can be realized by using materials with low atomic number for which the product $X_0\left\vert dE_{\mu}/dz \right\vert$ is maximized. 
The performance of a cooling cell can be characterized through the equilibrium emittance, which is obtained by setting $d\varepsilon_{\perp}/dz = 0$ and is given by
\begin{equation}\label{eq:eqmemit}
    \varepsilon_{\perp}^{eqm} \simeq \frac{\beta_{\perp} (13.6\,\textrm{MeV/c})^2}{2\beta m_{\mu}X_{0}} \left\vert\frac{dE_{\mu}}{dz}\right\vert^{-1}.
\end{equation}
Beams having emittances below equilibrium are heated while those having emittances above are cooled.

\section*{Experimental apparatus}\label{setup}

\begin{figure}[hbt!]%
\centering
\includegraphics[width=1.0\textwidth]{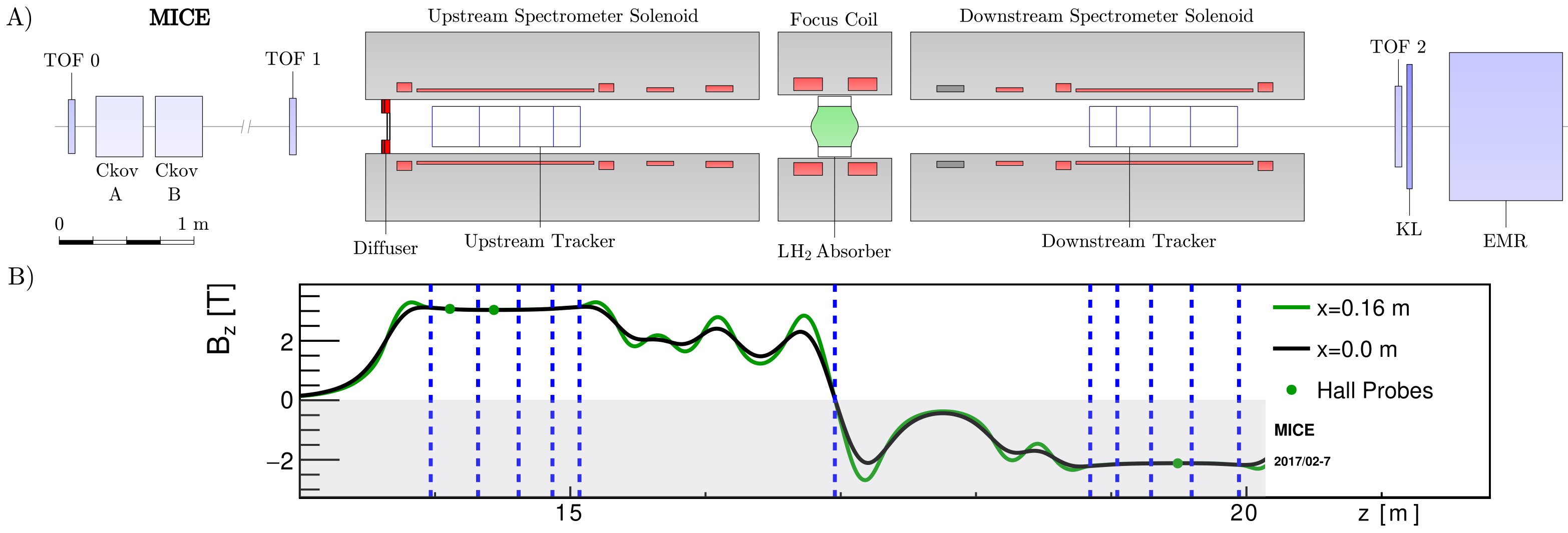}
\caption{a) Schematic layout of the MICE experimental setup and the modelled on-axis longitudinal magnetic field. Magnet coils are shown in red, the absorber in green and the various detectors are
individually labelled (see text for description). b) The modelled longitudinal magnetic field along the length of the MICE channel is shown on-axis (black line) and at 160 mm from the beam axis (green line) in the horizontal plane. The measurements of Hall probes situated at 160 mm from the axis are also shown (green circles). Vertical dashed lines indicate the positions of the tracker stations and the absorber.}\label{cc}
\end{figure}

The main component of the experiment was the MICE channel, a magnetic lattice of 12 strong-focusing superconducting coils symmetrically placed upstream and downstream of the absorber module. The MICE channel and instrumentation are shown schematically in Fig. \ref{cc}.

Muons were produced by protons from the ISIS synchrotron \cite{thomason2019isis} impinging on a titanium target \cite{booth2013design} and were delivered to the cooling channel via a transfer line \cite{bogomilov2012mice, adams2016pion}. Tuning the fields of two bending magnets in the transfer line enabled the selection of a beam with average momentum in the range 140--240$\, \mathrm{MeV/c}$. A variable-thickness brass and tungsten diffuser mounted at the entrance of the channel allowed the generation of beams with input emittance in the range 3--10$\,$mm.

The superconducting coils were grouped in three modules: two identical spectrometer solenoids situated upstream and downstream of the focus coil module that housed the absorber. Each spectrometer solenoid contained three coils that provided a uniform magnetic field of up to 4$\,$T in the tracking region, and two coils used to match the beam into or out of the focus-coil module. The focus-coil module contained a pair of coils designed to focus the beam tightly at the absorber. The large angular divergence (small $\beta_\perp$) of the focused beam reduced the emittance growth caused by multiple scattering in the absorber and increased the cooling performance. The two focus coils could be operated with identical or opposing magnetic polarities. For this study, the focus coils and the spectrometer solenoids were powered with opposite-polarity currents, thus producing a field that flipped polarity at the centre of the absorber. This magnetic field configuration was used to prevent the growth of beam canonical angular momentum. The field within the tracking regions was monitored using calibrated Hall probes. A soft-iron partial return yoke was installed around the magnetic lattice to contain the field.

Due to a magnet power lead failure during the commissioning phase, one of the matching coils in the downstream spectrometer solenoid was rendered inoperable. The built-in flexibility of the magnetic lattice allowed a compromise between the cooling performance and transmission that ensured the realisation of an unambiguous ionization cooling signal. 

As discussed above, absorber materials with low atomic numbers are preferred for ionization cooling lattices. Lithium hydride (LiH) and liquid hydrogen (LH$_2$) were the materials of choice in MICE. The lithium hydride absorber was a disk with a thickness of 65.37 $\pm$ 0.02$\,$mm and a density of \mbox{0.6957 $\pm$ 0.0006$\mathrm{\,g\,cm^{-3}}$} \cite{cooldemo}. The lithium used to produce the absorber had an isotopic composition of 95.52$\%$ $\ce{^{6}_{}Li}$ and 4.48$\%$ $\ce{^{7}_{}Li}$.

The liquid hydrogen was contained within a 22-litre aluminium vessel: a \mbox{300$\,$mm} diameter cylinder with a pair of dome-shaped containment windows at its ends \cite{bayliss2018liquid}. An additional pair of aluminium windows were mounted for safety purposes. The on-axis thickness of the liquid-hydrogen volume was \mbox{349.6 $\pm$ 0.2\,mm}. The density of the liquid hydrogen was measured to be 0.07053 $\pm$ 0.00008$\mathrm{\,g\,cm^{-3}}$ at 20.51$\,\mathrm{K}$ \cite{bogomilov2021performance}. The cumulative on-axis thickness of the aluminium windows was 0.79 $\pm$ 0.01$\,$mm. 

A comprehensive set of detectors were used to measure the particle species, position and momentum upstream and downstream of the absorber \cite{bogomilov2011mice,bogomilov2021performance}. The rate of muons delivered to the experiment was sufficiently low to allow the individual measurement of each incident particle. The data collected in cycles of several hours were aggregated offline and the phase-space occupied by the beam before and after the absorber was reconstructed. 

Upstream of the cooling channel, a velocity measurement provided by a pair of time-of-flight (TOF) detectors \cite{bertoni2010design} was used for electron and pion rejection. A pair of threshold Cherenkov counters \cite{cremaldi2009cherenkov} were used to validate the TOF measurement. Downstream, a further time-of-flight detector (TOF2) \cite{tof2}, a pre-shower sampling calorimeter (KL), and a fully active tracking calorimeter, the Electron Muon Ranger (EMR) \cite{asfandiyarov2013totally, adams2015electron}, were employed to identify electrons from muon decays that occurred within the channel and to validate the particle measurement and identification by the upstream instrumentation. Particle position and momentum measurements upstream and downstream of the absorber were provided by two identical scintillating fiber trackers \cite{ellis2011design} immersed in the uniform magnetic fields of the spectrometer solenoids.

Each tracker (named TKU and TKD, for upstream and downstream, respectively) consisted of five detector stations with a circular active area of 150$\,$mm radius. Each station comprised three planes of 350$\,\mu$m diameter scintillating
fibres, each rotated 120\degree\,with respect to its neighbour. In each station, the particle position was inferred from a coincidence of fiber signals. The particle momentum was reconstructed by fitting a helical trajectory to the reconstructed positions and accounting for multiple scattering and energy loss in the five stations \cite{dobbs2016reconstruction}. For particles with a helix radius comparable with the spatial kick induced by multiple scattering, the momentum resolution was improved by combining the tracker momentum measurement with the velocity measurement provided by the upstream TOF detectors. The measurements recorded by the tracker reference planes, at the stations closest to the absorber, were used to estimate beam emittance.

\section*{Observation of emittance reduction}\label{measurement}  

The data studied here were collected using beams that passed through a lithium hydride or a liquid hydrogen absorber. Scenarios with no absorber present or the empty liquid hydrogen vessel were also studied for comparison. For each absorber setting, three beam-line configurations were used to deliver muon beams with nominal emittances of 4, 6 and 10$\,$mm and a central momentum of approximately 140$\,$$\mathrm{MeV/c}$ in the upstream tracker. 
For each beam-line/absorber configuration, the final sample contained particles that were identified as muons by the upstream TOF detectors and tracker and had one valid reconstructed trajectory in each tracker. The kinematic, fiducial and quality selection criteria for the reconstructed tracks are listed in the Methods section. A Monte Carlo simulation of the whole experiment was used to estimate the expected cooling performance and to study the performance of the individual detectors \cite{maus2019}.

\begin{figure}[hbt!]%
\centering
\includegraphics[width=0.8\textwidth]{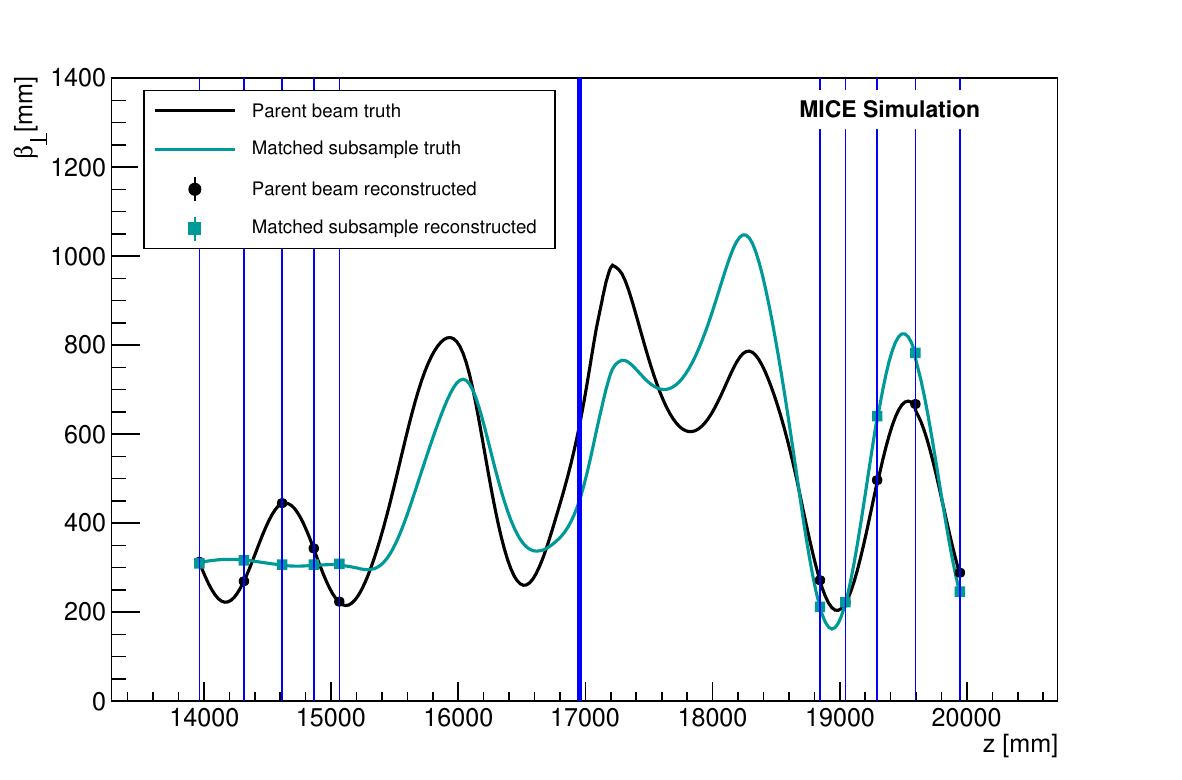}
\caption{Simulated evolution of the transverse betatron function, $\beta_\perp$, through the cooling channel containing the full liquid hydrogen vessel for the (black) parent beam and the (dark cyan) matched subsample. The corresponding lines represent the simulation truth, while the circles and squares at the (vertical blue lines) tracker stations represent the reconstructed simulation. The thick vertical blue line marks the central position of the absorber.
}\label{fig:beta}
\end{figure}

The beam matching into the channel slightly differed from the design beam optics due to inadequate focusing in the final section of the transfer line. This mismatch resulted in an oscillatory behaviour of the transverse betatron function in the upstream tracker region and an increased, sub-optimal $\beta_\perp$ at the absorber, which degraded the cooling performance. An algorithm based on rejection sampling was developed to select beams with a constant betatron function in the upstream tracker, in agreement with the design beam optics. The selection was performed on the beam ensemble measured in the upstream tracker and was enabled by the unique MICE capability to measure muon beams particle by particle. An example comparison between the betatron function of an unmatched parent beam and that of a matched subsample is shown in Fig. \ref{fig:beta}. The $\beta_\perp$ of the subsample is approximately constant in the upstream tracker and, as a consequence, its value at the absorber centre is $\sim$$\,$28\% smaller than the corresponding value of the parent beam.

The sampling algorithm enabled the selection of subsamples with specific emittances. This feature was exploited to study the dependence of the cooling effect on input emittance. For each absorber setting, the three parent beams were each split into two distinct samples and six statistically independent beams with matched betatron functions ($\beta_\perp = 311$$\,$mm, $d\beta_\perp/dz$ = 0) and emittances of 1.5, 2.5, 3.5, 4.5, 5.5 and 6.5$\,$mm at the upstream tracker were sampled. The numbers of muons in each sample are listed in Extended Data Table \ref{tab:final_numbers}. The two-dimensional projections of the phase-space of the sampled beams on the transverse position and momentum planes are shown in Extended Data Fig. \ref{fig:position} and Extended Data Fig. \ref{fig:momentum}, respectively.   

\begin{figure}[hbt!]%
\centering
\includegraphics[width=0.64\textwidth]{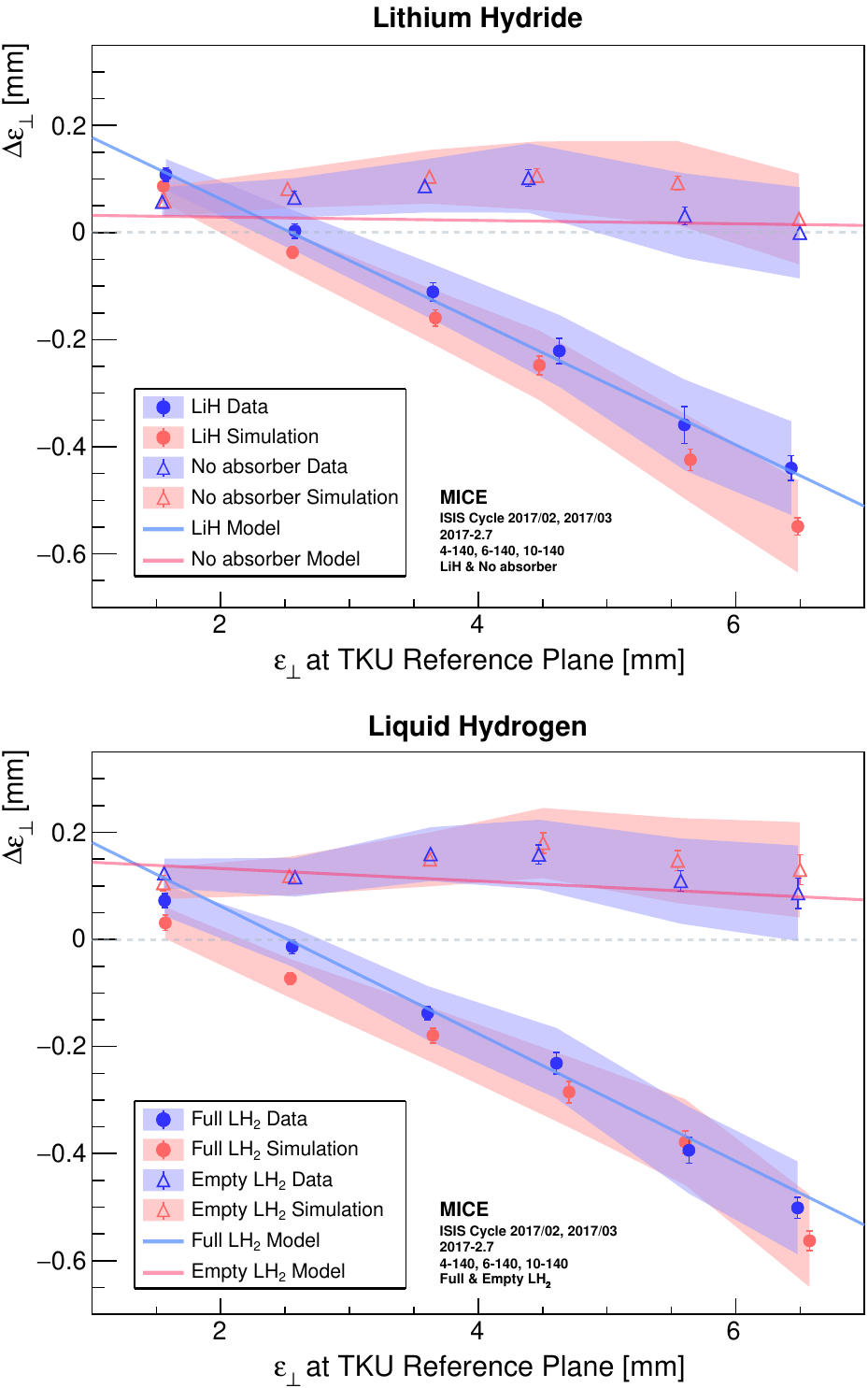}
\caption{Emittance change between the upstream and
 the downstream tracker reference planes as a function of emittance at the upstream tracker (TKU), for 140 $\mathrm{MeV/c}$ beams crossing (top) the lithium hydride and (bottom) the liquid hydrogen MICE absorbers. Results for the empty cases, `No absorber' and `Empty LH$_2$', are also shown. The measured effect is shown in blue, while the simulation is shown in red. The statistical uncertainty is indicated by the error bars, while the total uncertainty is shown by the semi-transparent fill. The solid lines represent the approximate theoretical model defined by equation \ref{eq:theo_cool} (see Methods section) for the (light blue) absorber and (light pink) empty cases.}
\label{fig:money}
\end{figure}

Fig. \ref{fig:money} shows the emittance change induced by the lithium hydride and the liquid hydrogen absorbers, as well as the corresponding empty cases, for each emittance subsample. The measurement uncertainty, depicted by the coloured bands, is dominated by systematic uncertainties. A correction was made to account for detector effects and for the inclusion only of events that reached the downstream tracker. Good agreement between data and simulation is observed in all configurations. The reconstructed data agree well with the model prediction. The model includes the heating effect in the aluminium windows (see Methods). 

The empty absorber cases show no cooling effects. In the empty channel case (`No absorber'), slight heating occurs due to optical aberrations and scattering in the aluminium windows of the two spectrometer solenoids. Additional heating caused by scattering in the liquid hydrogen vessel windows is observed in the `Empty LH$_{2}$' case. The `LiH' and `Full LH$_{2}$' absorbers demonstrate emittance reduction for beams with emittances larger than $\sim$$\,$2.5$\,$mm. This is a clear signal of ionization cooling, a direct consequence of the presence of absorber material in the path of the beam.

For beams with 140$\,$$\mathrm{MeV/c}$ momentum and $\beta_\perp$ = 450$\,$mm at the absorber, the theoretical equilibrium emittances of the MICE lithium hydride and liquid hydrogen absorbers, including the contributions from the corresponding set of aluminium windows, are $\sim$$\,$2.5$\,$mm in both cases. By performing a linear fit to the measured cooling trends in Fig. \ref{fig:money}, the effective equilibrium emittances of the absorber modules are estimated to be 2.6$\,$$\pm$$\,$0.4$\,$mm for lithium hydride and 2.4$\,$$\pm$$\,$0.4$\,$mm for liquid hydrogen. The parameters of the linear fits to the four emittance change trends are shown in Table \ref{tab:grad_int}. Our null hypothesis was that for each set 
of six input beam settings, the slopes of the emittance change trends in the presence and absence of an absorber are compatible. A Student's $t$-test found that the probabilities of observing the effects measured here are lower than $10^{-5}$ for both the `LiH'--`No absorber' and `Full LH$_{2}$'--`Empty LH$_{2}$' pairs, hence the
null hypotheses were rejected.

\begin{table}[hbt!]
\begin{center}
\caption{Parameters of the linear fits performed on the emittance change trends corresponding to the four absorber configurations.}
\begin{tabular}{p{3.3cm}|c|c}

 Absorber Configuration & Intercept [mm] & Slope\\
 \hline
 No absorber    & 0.102 $\pm$ 0.007 & -0.011 $\pm$ 0.012\\
 LiH            & 0.297 $\pm$ 0.006 & -0.115 $\pm$ 0.013\\
 Empty LH$_2$   & 0.150 $\pm$ 0.005 & -0.006 $\pm$ 0.013\\
 Full LH$_2$    & 0.279 $\pm$ 0.007 & -0.118 $\pm$ 0.013\\

 
\end{tabular}
\label{tab:grad_int}
\end{center}    
\end{table}

\section*{Conclusions}

The results reported here represent the first direct measurement of normalized transverse emittance reduction of a muon beam by ionization cooling. They demonstrate the viability of this novel cooling technique as a means of producing low-emittance muon beams. The demonstration of ionization cooling by the MICE collaboration constitutes a substantial and encouraging breakthrough in the R\&D efforts to deliver high-brightness muon beams suitable for high-intensity muon-based facilities such as a muon collider or a neutrino factory.

\newpage

\section*{Methods}\label{sec:methods}


\subsection*{Event reconstruction}\label{eventrec}

Each TOF hodoscope was composed of two planes of scintillator slabs oriented along the $x$ and $y$ directions. Photomultiplier tubes (PMTs) at both ends of each slab were used to collect and amplify the signal produced by a charged particle traversing the slab. A coincidence of signals from the PMTs of a slab was recorded as a `slab hit'. A pair of orthogonal slab hits formed a space point. The information collected by the four corresponding PMTs was used to reconstruct the position and the time at which the particle passed through the detector. For a detailed description of the TOF time calibration see \cite{calib}. The MICE data acquisition system readout was triggered by a coincidence of signals from the PMTs of a single slab of the TOF1 detector. All the data collected by the detector system after each TOF1 trigger were aggregated and formed a particle event.

For each tracker, signals from the scintillating fibres in the five stations were combined to reconstruct the helical trajectories of the traversing charged particles. The quality of each fitted track was indicated by the $\chi^2$ per degree of freedom

\begin{equation}
    \chi^2_{\mathrm{dof}} = \frac{1}{n-5} \sum\limits_{i=1}^{n} \frac{\delta x^2_i}{\sigma^2_i},
\end{equation}
where $n$ is the number of tracker planes that contributed to the reconstruction, $\delta x_i$ is the distance between the measured position in the $i$th tracker plane and the fitted track and $\sigma_i$ is the position measurement resolution in the tracker planes. A more detailed description of the reconstruction procedure and its performance can be found in \cite{bogomilov2021performance,dobbs2016reconstruction}.

\subsection*{Sample selection}\label{eventselect}

The measurements taken by the detector system were used to select the final sample. The following selection criteria ensured that a pure muon beam, with a narrow momentum spread, and fully transmitted through the channel, was selected for analysis:

\begin{itemize}

  \item One reconstructed space point found in TOF0 and TOF1, and one reconstructed track found in TKU and TKD;
  \item Time-of-flight between TOF0 and TOF1 consistent with that of a muon;
  \item Momentum measured in TKU consistent with that of a muon, given the TOF0-TOF1 time-of-flight;
  \item In each tracker, a reconstructed track contained within the cylindrical fiducial volume defined by a radius of 150$\,$mm and with $\chi^2_{\mathrm{dof}} < 8$;
  \item Momentum measured in TKU in the 135--145$\,$$\mathrm{MeV/c}$ range;
  \item Momentum measured in TKD in the 120--170$\,$$\mathrm{MeV/c}$ range for the empty absorber configurations, and 90--170$\,$$\mathrm{MeV/c}$ range for the LiH and LH$_2$ absorbers;
  \item At the diffuser, a track radial excursion contained within the diffuser aperture radius by at least 10$\,$mm.
\end{itemize}
The same set of selection criteria was applied to the simulated beams.

\subsection*{Beam sampling}\label{sampling}

The sampling procedure developed to obtain beams matched to the upstream tracker is based on a rejection sampling algorithm \cite{von195113,wells2004generalized}. It was designed to carve out a beam subsample that followed a four-dimensional Gaussian distribution described by a specific (target) covariance matrix from an input beam ensemble (parent). 

The custom algorithm required an estimate of the probability density function underlying the beam ensemble. Since the MICE beams were only approximately Gaussian and approximately cylindrically symmetric, the kernel density estimation (KDE) technique was used to evaluate the parent beam density in a non-parametric fashion \cite{silverman2018density,scott2015multivariate}. In KDE, each data point is assigned a smooth weight function, also known as the kernel, and the contributions from all data points in the data set are summed. The multivariate kernel density estimator at an arbitrary point $\boldsymbol{\mathrm{u}}$ in the $d$-dimensional space is given by

\begin{equation}
    \hat{f}(\boldsymbol{\mathrm{u}}) = \frac{1}{nh^d}\sum\limits_{i=1}^{n} K\left(\frac{\boldsymbol{\mathrm{u}}-\boldsymbol{\mathrm{u}}_\mathrm{i}}{h}\right),
    \label{eq:kde}
\end{equation}
where $K$ is the kernel, $n$ the sample size, $h$ the width of the kernel, and $\boldsymbol{\mathrm{u}}_\mathrm{i}$ represents the coordinate of the $i$-th data point in the sample. In this analysis, Gaussian kernels of the following form were used 

\begin{equation}
    K\left(\frac{\boldsymbol{\mathrm{u}}-\boldsymbol{\mathrm{u}}_\mathrm{i}}{h}\right) = \frac{1}{\sqrt{(2\pi)^d\left\vert\boldsymbol{\Sigma}_\perp\right\vert}}\exp\left[-\frac{1}{2}\frac{(\boldsymbol{\mathrm{u}}-\boldsymbol{\mathrm{u}}_\mathrm{i})^{T}\boldsymbol{\Sigma}^{-1}_\perp(\boldsymbol{\mathrm{u}}-\boldsymbol{\mathrm{u}}_\mathrm{i})}{h^2}\right],
    \label{eq:kdekernel}
\end{equation}
where $\boldsymbol{\Sigma}_\perp$ is the covariance matrix of the data set. The width of the kernel is chosen to minimise the mean integrated squared error (MISE), which measures the accuracy of the estimator \cite{marron1992exact}. Scott's rule of thumb was followed in this work, wherein the kernel width was determined from the sample size $n$ and the number of dimensions $d$ through $h = n^{-1/(d+4)}$ \cite{scott2015multivariate}. 

The KDE form described in equations \ref{eq:kde} and \ref{eq:kdekernel} was used to estimate the transverse phase-space density of the initial, unmatched beams, with the estimated underlying density denoted by $Parent(\boldsymbol{\mathrm{u}})$. The target distribution, $Target(\boldsymbol{\mathrm{u}})$, is a 4-D Gaussian defined by a covariance matrix parameterised through the transverse emittance ($\varepsilon_\perp$), transverse betatron function ($\beta_\perp$), mean longitudinal momentum and mean kinetic angular momentum \cite{penn2000beam}.

The sampling was performed on the beam ensemble measured at the upstream tracker station closest to the absorber. For each particle in the parent beam, with 4-D phase-space vector $\boldsymbol{\mathrm{u}}_\mathrm{i}$, the sampling algorithm worked as follows:

\begin{enumerate}
 
    \item Compute the selection probability as 
    \begin{equation}
     P_{select}(\boldsymbol{\mathrm{u}}_\mathrm{i}) = \mathcal{C}\times \frac{Target(\boldsymbol{\mathrm{u}}_\mathrm{i})}{Parent(\boldsymbol{\mathrm{u}}_\mathrm{i})},    
    \end{equation}
     where the normalisation constant $\mathcal{C}$ ensures that the selection probability $P_{select}(\boldsymbol{\mathrm{u}}_\mathrm{i}) \leq 1$;
     
     \item Generate a number $\xi_i$ from the uniform distribution $\mathcal{U}([0,1])$;
     \item If $P_{select}(\boldsymbol{\mathrm{u}}_\mathrm{i}) > \xi_i$, then accept the particle. Otherwise, reject it.
     
\end{enumerate}
The normalisation constant $\mathcal{C}$ was calculated prior to the sampling iteration presented in steps 1--3. It required an iteration through the parent ensemble (of size $n$) and it was calculated as

\begin{equation}
    \mathcal{C} = \min_{i\in\{1, ..., n\}}\frac{Parent(\boldsymbol{\mathrm{u}}_\mathrm{i})}{Target(\boldsymbol{\mathrm{u}}_\mathrm{i})}.
\end{equation}

The target parameters of interest were $\varepsilon_\perp$, $\beta_\perp$ and $\alpha_\perp = - \frac{1}{2} d\beta_\perp/dz$. For beams with central momentum of 140$\,\mathrm{MeV/c}$ and a solenoidal magnetic field of 3$\,$T, the matching conditions in the upstream tracker were \mbox{($\beta_\perp$, $\alpha_\perp$) = (311$\,$mm, 0)}. The target mean kinetic angular momentum was kept at the value measured in the parent beam for which the sampling efficiency was at a maximum.

\subsection*{Emittance change calculation and model}\label{model}

The emittance change measured by the pair of MICE scintillating fibre trackers is defined as

\begin{equation}
    \Delta\varepsilon_\perp = \varepsilon_{\perp}^{d} - \varepsilon_{\perp}^{u},
\end{equation}
where $\varepsilon_{\perp}^{d}$ is the emittance measured in the downstream tracker and $\varepsilon_{\perp}^{u}$ is the emittance measured in the upstream tracker. In each tracker the measurement is performed at the station closest to the absorber.

Starting from the cooling equation shown in Eq. \ref{eq:cooleqn}, the emittance change induced by an absorber material of thickness $z$ can be expressed as a function of the input emittance, $\varepsilon_{\perp}^{u}$, as follows:

\begin{equation}
    \Delta\varepsilon_\perp(\varepsilon_{\perp}^{u}) \approx (\varepsilon_{\perp}^{eqm} - \varepsilon_{\perp}^{u})\left[1-\exp\left(-\frac{\vert dE_{\mu}/dz\vert}{\beta^2E_{\mu}}z\right)\right],
    \label{eq:theo_cool}
\end{equation}
where $\varepsilon_{\perp}^{eqm}$ is the equilibrium emittance and the mean energy loss rate, $\vert dE_{\mu}/dz\vert$, is described by the Bethe-Bloch formula \cite{leroy2011principles}.

The expected emittance change depends on the type and amount of material that the beam traverses between the two measurement locations. Aside from the absorber material under study and absorber module windows, the beam crossed an additional pair of aluminium windows, one downstream of TKU, and the other upstream of TKD. All windows were made from Al 6061-T651 alloy. Equation \ref{eq:theo_cool} was used to estimate the theoretical cooling performance, including the effect of aluminium windows. The properties of the absorber and window materials required for the calculation are shown in Table \ref{tab:materials}. For each absorber configuration, the beam properties required for the model ($\beta, \beta_{\perp}, E_{\mu}$) were obtained from the simulation of the 3.5$\,$mm beam.

\begin{table}[hbt!]
    \centering
        \caption{Material properties of the MICE lithium hydride and liquid hydrogen absorbers, as well as those of the aluminium alloy used for the windows \cite{particle2020review}. $Z$ and $A$ are the atomic and mass numbers of the material, respectively, and $I$ is the mean excitation energy of the atoms in the material.}
\begin{tabular}{p{3.3cm}|c|c|c}

 \diagbox[innerwidth=3.3cm]{Property}{Material} & MICE LiH & Liquid H$_2$ & Al 6061-T651\\
 \hline
 Density, $\rho$ [g/cm$^3$]   & 0.6957 & 0.07053 & 2.727\\
 $\langle Z/A\rangle$         & 0.56716 & 0.99212 & 0.48145\\
 $I$ [eV]                       & 36.5 & 21.8 & 166\\
 $X_{0}$ [cm]                 & 102.04 & 866 & 8.68\\
 
\end{tabular}
    \label{tab:materials}
\end{table}

\clearpage

\backmatter

\section*{Data availability}

The unprocessed and reconstructed data that support the findings
of this study are publicly available on the GridPP computing Grid at \href{https://doi.org/10.17633/rd.brunel.3179644}{https://doi.org/10.17633/rd.brunel.3179644} (MICE unprocessed data) and \href{https://doi.org/10.17633/rd.brunel.5955850}{https://doi.org/10.17633/rd.brunel.5955850} (MICE reconstructed data).

Publications using MICE data must contain the following statement:
“We gratefully acknowledge the MICE collaboration for allowing us
access to their data. Third-party results are not endorsed by the MICE
collaboration.”

\section*{Code availability}

The MAUS software that was used to reconstruct and analyse the
MICE data is available at \href{https://doi.org/10.17633/rd.brunel.8337542}{https://doi.org/10.17633/rd.brunel.8337542}. The analysis presented here used MAUS version 3.3.2.

\bmhead{Acknowledgments}

The work described here was made possible by grants from the Science
and Technology Facilities Council (UK), the Department of Energy and the National Science
Foundation (USA), the Istituto Nazionale di Fisica Nucleare (Italy), the European Union under the European Union’s Framework Programme 7 (AIDA project, grant agreement number
262025; TIARA project, grant agreement number 261905; and EuCARD), the Japan Society for
the Promotion of Science, the National Research Foundation of Korea (number NRF2016R1A5A1013277), the Ministry of Education, Science and Technological Development of the Republic of Serbia, the Institute of High Energy Physics/Chinese Academy of Sciences fund for collaboration between the People’s Republic of China and the USA, and the Swiss National Science Foundation in the framework of the SCOPES programme. We gratefully acknowledge all sources of support. We are grateful for the support given to us by the staff of the STFC Rutherford Appleton and Daresbury laboratories. We acknowledge the use of Grid computing resources deployed and operated by GridPP in the UK, \href{http://www.gridpp.ac.uk/}{http://www.gridpp.ac.uk/}. This publication is dedicated to the memory of Vittorio Palladino and Don Summers who passed away while the data analysis from which the results presented here was being developed.

\bmhead{Author contributions}

All authors contributed considerably to the design or construction of the
apparatus or to the data taking or analysis described here.

\bmhead{Competing interests}

The authors declare no competing interests.

\begin{appendices}
\renewcommand\theHtable{AABB\arabic{table}}   
\renewcommand\theHfigure{AABB\arabic{figure}}

\newpage
\section{Extended Data}\label{exdata}

\begin{table}[hbt!]
\begin{center}
\caption{The sample size of the matched beams.}
\begin{tabular}{p{4.3cm}|c|c|c|c|c|c}

 \diagbox[innerwidth=4.3cm]{Absorber}{Input $\varepsilon_\perp$ [mm]} & 1.5 & 2.5 & 3.5 & 4.5 & 5.5 & 6.5\\
 \hline
 Empty LH$_2$   & 8141 & 10162 & 19525 & 29896 & 13196 & 22080\\
 Full LH$_2$    & 5199 & 8541 & 16757 & 20836 & 9063 & 15326\\
 No absorber    & 4496 & 4792 & 32836 & 17659 & 5324 & 8573\\
 LiH            & 4549 & 4372 & 9150 & 21071 & 3927 & 7618\\
 
\end{tabular}
\label{tab:final_numbers}
\end{center}    
\end{table}

\clearpage
\newpage

\begin{figure}[p]%
\centering
\includegraphics[width=1.0\textwidth]{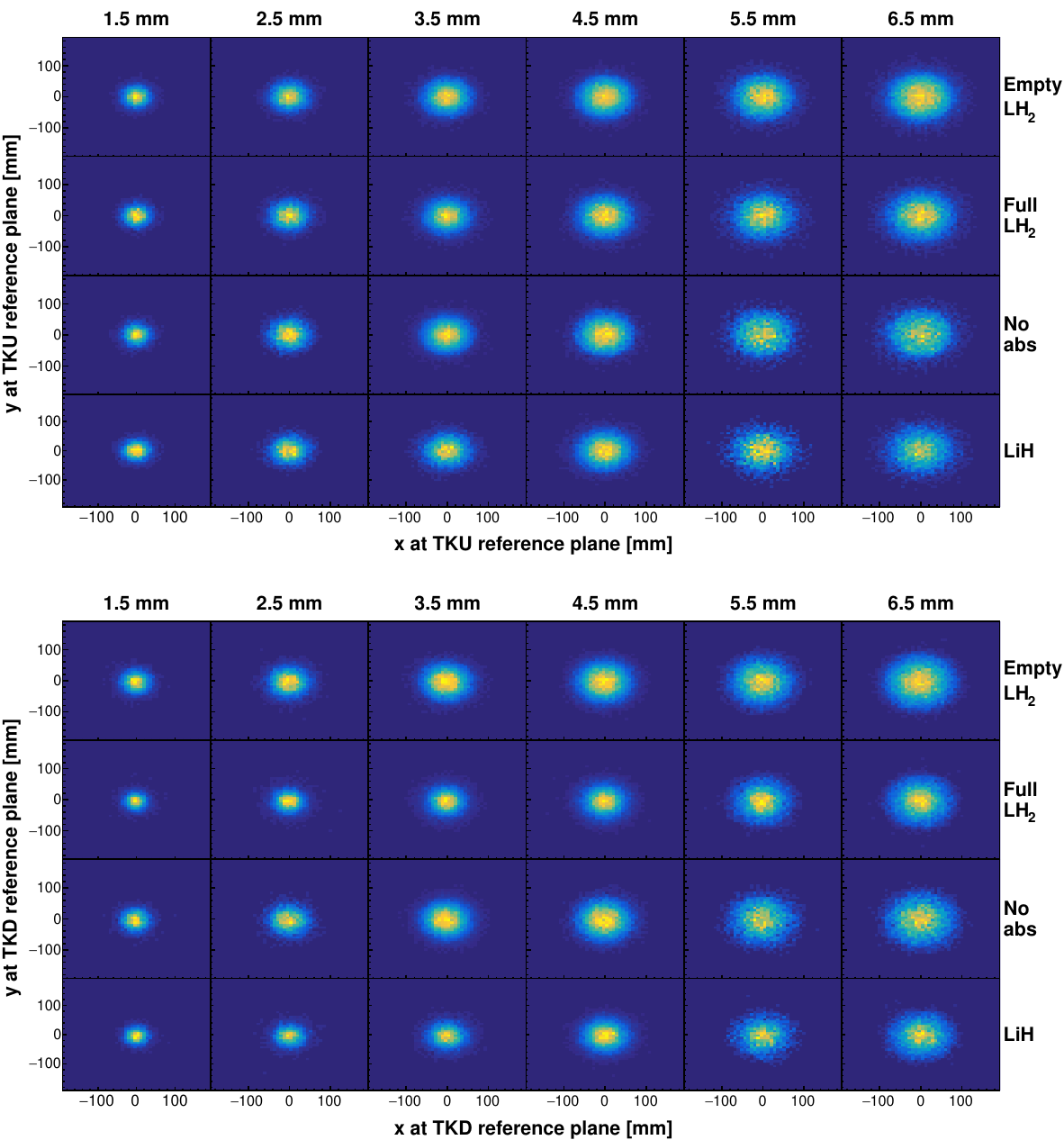}
\caption{Transverse beam profiles in the (top) upstream and (bottom) downstream trackers. The beams that pass through an absorber present a smaller transverse size in the downstream tracker than the beams that traverse an empty absorber module. This effect is caused by a change in focusing due to energy loss.}
\label{fig:position}
\end{figure}

\begin{figure}[hbt!]%
\centering
\includegraphics[width=1.0\textwidth]{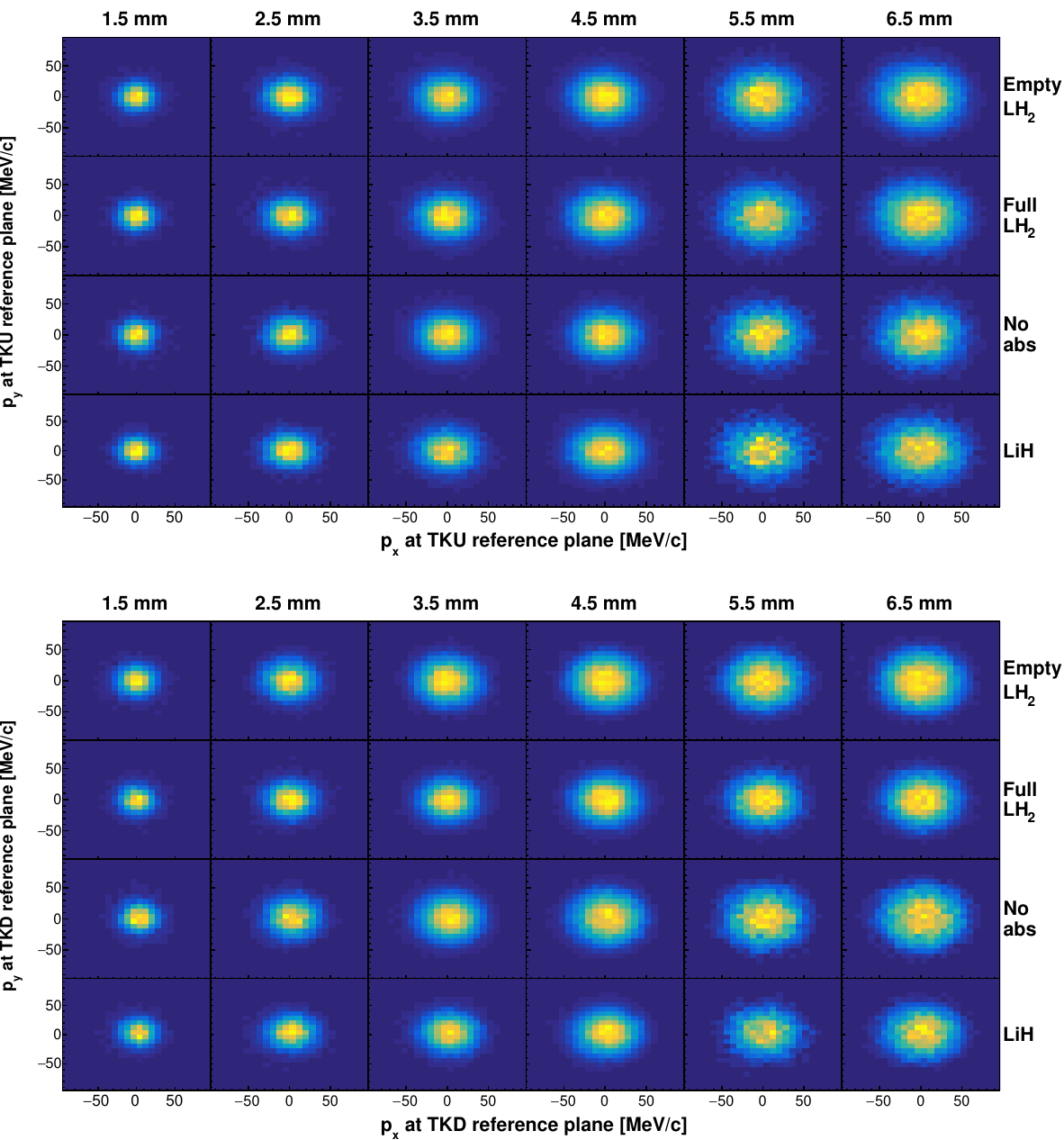}
\caption{Transverse beam momentum in the (top) upstream and (bottom) downstream trackers. The beams that pass through an absorber present a smaller transverse momentum in the downstream tracker than the beams that traverse an empty absorber module.}
\label{fig:momentum}
\end{figure}

\end{appendices}
\clearpage
\newpage

\bibliography{sn-bibliography}

\end{document}